\documentclass[12pt]{article}
\usepackage{cite}
\usepackage{amsbsy}
              
\title{Geometries of Maximal Acceleration.}
\author{M. Toller \thanks{e-mail: toller@iol.it}\\ 
via Malfatti n. 8  \\
I-38100 Trento, Italy}
 
\begin{document} 
\maketitle
                 
\begin{abstract}
We discuss and compare several geometric structures which imply an upper bound to the acceleration of a particle measured in its rest system. While all of them have the same implications on the motion of a point particle, they differ in other important respects. In particular, they have different symmetry groups, which influence in a different way the search for an underlying dynamical theory.


11.30.Cp  (Lorentz and Poincar\'e invariance);

02.40.Ft  (Convex sets and geometric inequalities).

\end{abstract}

\newpage

\section{Introduction.}

Starting from a letter published by Caianiello in 1981 \cite{Caianiello1}, several ideas have been proposed concerning a possible upper bound to the acceleration of a particle \cite{Toller1,CDFMV,Brandt1,Brandt2,Scarpetta,Caianiello2,CL,SS,GS,Brandt3,Toller2,Brandt4,WPI1,WPI2,Brandt5,NFS,Rama,Castro,Papini}. More recently, further important developments have appeared \cite{Schuller,Schuller1,SWG,SP,Schuller2}. For lack of space, we don't cite here many other papers devoted to applications of the maximal acceleration idea.

There are two points of view, which are physically rather different, though many formal considerations can be applied to both. Some authors \cite{Brandt1,Brandt2,Scarpetta,CL,Brandt3,Brandt4} assume a maximal acceleration valid for all the kinds of particles. If it has to be explained in terms of the known general physical principles, from a dimensional argument we find
\begin{equation}
a_M \approx (G \hbar)^{-1/2} c^{7/2},
\end{equation}
where $a_M$ is the maximal acceleration in the rest frame and $G$ is the gravitational coupling constant. The constants appearing in this formula suggest that an explanation should be based on a relativistic quantum theory of gravitation.

Other authors \cite{Caianiello2,SS,WPI1,Papini} propose an upper bound depending on the mass $m$ of the particle. If we assume that gravitational phenomena are irrelevant, from a dimensional argument we find
\begin{equation}
a_M \approx m \hbar^{-1} c^3.
\end{equation}
The same assumption can equally well be described as a mass-dependent upper bound to the force (always measured in the rest system) given by 
\begin{equation}
f_M \approx m^2 \hbar^{-1} c^3.
\end{equation}

These formulas suggest that one should explain these upper bounds in terms of a special-relativistic quantum theory and many arguments in this direction have been proposed. However, in our opinion, no formal rigorous proof has been given up to now. Presumably, a rigorous treatment should involve the locality principle, namely it should proceed in the framework of quantum field theory. Then, during an interaction the production of new particles is inavoidable and one may think that, when the acceleration is too large, the system cannot be considered any more as a single particle state. In this way, however, it is difficult to find a sharp limitation.

The aim of the present article is not to give a proof, or at least a new justification, of the maximal acceleration hypothesis, but to discuss its formulations in terms of geometric concepts.  We think that this discussion is relevant, since various authors use rather different geometric constructions, and the connection between them is not trivial.

We use as a guide the familiar geometric description of the maximal velocity which appears in relativity theory. It is given in terms of the closed future cone $V^+$ in the Minkowski space-time. In the presence of gravitation, one has to consider a cone in each tangent space of a pseudo-Riemannian space-time $\mathcal{M}$.  The vectors tangent to the world-line of a particle must belong to these cones.

In the following a {\it wedge} means a convex dilatation-invariant subset of a vector space and a {\it cone} is a wedge that does not contain straight lines. According to this definition, a cone is necessarily convex. The set $V^+$ has the following properties:
\renewcommand{\labelenumi}{\alph{enumi})}
\begin{enumerate}
\item It is a closed cone with nonempty interior;
\item It is invariant under the rotation group.
\end{enumerate}
One can easily show (see the appendix A) that these conditions determine $V^+$ completely up to the choice of the numerical value $c$ of the light velocity and of the direction of time. 

We see that the convexity requirement is very powerful and it is physically well justified. It means that if a particle moves during two consecutive time intervals with two allowed constant velocities, the overall average velocity is also allowed.

We see that a cone satisfying the conditions a), b) listed above is necessarily symmetric with respect to the orthochronous Lorentz group $O^{\uparrow}(1, 3)$ and one may consider this argument as a derivation of the Lorentz symmetry from the rotation symmetry and the existence of a maximal velocity. The next step towards a relativistic theory is to assume that the Lorentz group is not only the symmetry group of $V^+$, but also the symmetry group of all the physical laws. This approach to special relativity is certainly less satisfactory than the usual one, based on the relativity principle, but it can be extended by analogy to a treatment of theories with a maximal acceleration.

In order to introduce a maximal acceleration, one has to consider a manifold with at least seven dimensions, which describes, besides time and position, the velocity (or the energy-momentum) of a particle. The maximal acceleration (or the maximal force) is described by cones in the tangent spaces. We shall show that in some cases, assuming the topological properties a) given above and Lorentz symmetry, one can determine these cones up to the choice of the numerical values of the light velocity $c$ and of the  maximal acceleration $a_M$. 

The cones obtained in this way have a symmetry group which acts linearly on the tangent spaces of the manifold considered (but not necessarily on the manifold itself). This group may be
larger than $SO^{\uparrow}(1, 3)$ and, following the example of relativity, one may try to assume that it (or one of its subgroups larger than $SO^{\uparrow}(1, 3)$) also represents a symmetry of the dynamical equations. Of course, this higher symmetry, which is not physically observed, has to be a broken symmetry, namely the vacuum is only symmetric with respect to the Lorentz subgroup.

In the following three sections we consider some different geometric schemes of the kind described above, based on different manifolds, and we find that they give rise to different enlarged symmetry groups. This means that there is some deep physical difference between these approaches. In Section 5 we introduce a different physical interpretation which permits a better understanding of the problem. In the last section we draw some conclusions and we indicate some possible future developments.  

\section{The space-time-velocity manifold.}

To the best of our knowledge, the first geometrical scheme which implies a maximal acceleration is given by Born's duality theory \cite{Born1,Born2}. It is based on the eight-dimensional relativistic phase space $\mathcal{W} = \mathbf{R}^8$ the coordinates of which are the space-time coordinates $x^k$ and the components $p^k$ of the energy-momentum four-vector. In the following, the indices $i, j, k$ take the values $0, 1, 2, 3$, we assume $c = 1$ and for the relativistic scalar product of two four-vectors we use the notation $x \cdot p = x^k p_k = x^0 p^0 - \mathbf{x} \cdot \mathbf{p}$. One introduces a  metric of the form
\begin{equation}
d\sigma^2 = dx \cdot dx + f_M^{-2} \, dp \cdot dp,
\end{equation}
where $f_M$ is the maximal force. Here $p$ represents the canonical four-momentum, which satisfies the canonical Poisson brakets. 

If we consider a free particle with mass $m$, we have $p = m u$, where $u$ is the four-velocity, and the same expression can be written in the form
\begin{equation} \label{Metric}
d\sigma^2 = dx \cdot dx + a_M^{-2} \, du \cdot du.
\end{equation}
Since we are concerned with the acceleration, we have to consider interacting particles and the two expressions given above are not equivalent any more. 

If a particle interacts with an electromagnetic field, the first expression is not gauge invariant and we prefer to concentrate our attention on the second expression and on a space $\mathcal{W} = \mathbf{R}^8$ with coordinates $x^k$ and $u^k$. It can be interpreted as the tangent bundle of the Minkowski space-time and many of the following considerations can be generalized to the tangent bundle of a pseudo-Riemannian space-time. 

One can also introduce the kinetic four-momentum, which by definition is given by $mu$. In general, its Poisson brackets are more complicated than the canonical ones. In the present article we deal with the kinematical aspects of the problem and we do not enter into the details of the Hamiltonian dynamics.

We describe the world line of a massive particle by means of the functions $\tau \to x(\tau)$, where the parameter $\tau$ is the proper time, defined by
\begin{equation}
d\tau^2 = dx \cdot dx.
\end{equation}
The curves in $\mathcal{W}$ which describe the motion of a particle must satisfy the equation
\begin{equation}
\frac{dx}{d\tau} = u,
\end{equation}
which implies the constraints
\begin{equation} \label{Constr1}
u \cdot u = 1, \qquad u^0 \geq 1,
\end{equation}
\begin{equation} \label{Constr2}
d \mathbf{x} = (u^0)^{-1} \mathbf{u} dx^0.
\end{equation}
An interesting consequence of them, considered in the next section, is
\begin{equation} \label{Constr3}
dx \cdot du = 0.
\end{equation}

The constraint (\ref{Constr1}) is {\it holonomous} and it defines  a 7-dimensional submanifold $\mathcal{V}$ of $\mathcal{W}$, which we call the {\it space-time-velocity manifold}. Only this submanifold is involved in the description of the motion of a particle. From the dynamical point of view, we are dealing with a constrained Hamiltonian system \cite{Dirac}. The manifold $\mathcal{V}$ has a pre-symplectic structure \cite{Souriau}, that determines the dynamics without any reference to the larger space $\mathcal{W}$. Borrowing a term from quantum field theory, one may say that $\mathcal{W}$ also describes ``off shell'' particles.

The constraint (\ref{Constr2}) is {\it anholonomous}, since it involves the tangent vectors. More correctly, rather than a constraint, it should be considered ad a dynamical equation, which might take a different form in a modified dynamical scheme. In section 5 we discuss a physical interpretation which goes beyond the particle kinematics and does not require the constraint (\ref{Constr2}), but only the holonomous constraint (\ref{Constr1}).

It is natural to impose the condition
\begin{equation} \label{Cond1}
d\sigma^2 \geq 0.
\end{equation}
Note that the expression (\ref{Metric}) and the condition (\ref{Cond1}) are invariant under the pseudo-orthogonal group $O(2, 6)$, which may be called the ``maximal acceleration group''  \cite{Brandt2}.  We shall find other groups that deserve this denomination equally well. 

If one restricts his attention to the canonical transformations, one is led to consider the intersection $O(2, 6) \cap Sp(4)$, which is isomorphic to the pseudo-unitary group $SU(1, 3)$ \cite{Brandt5,Low1,Low2,Low3,Rama,Castro}.  One has to be careful when dealing with interacting particles, since the symplectic group $Sp(4)$ acts on the canonical four-momentum, while the pseudo-orthogonal group $O(2, 6)$ acts on the four-velocity or on the kinetic four-momentum. 

It is important to stress that the constraints (\ref{Constr1}) and (\ref{Constr2}) are not invariant under the group $O(2, 6)$, which does not transform a particle world line into another one. In other words, the dynamics, as we know it, has not this symmetry. The subset of a tangent space $W$ of $\mathcal{W}$ defined by eq.\ (\ref{Cond1}) is invariant under the dilatation group $\mathbf{R}^*$, but it is not convex and it is not a cone.

If we take into account the holonomous constraint (\ref{Constr1}), we have
\begin{equation}
du^0 = (u^0)^{-1} \mathbf{u} \cdot d \mathbf{u}
\end{equation}
and the condition (\ref{Cond1}) takes the form
\begin{equation} \label{Cond2}
\|d \mathbf{x}\|^2 + a_M^{-2} \left(\|d\mathbf{u}\|^2 - (u^0)^{-2} (\mathbf{u} \cdot d \mathbf{u})^2\right) \leq (dx^0)^2.
\end{equation}

If we add the condition $dx^0 \geq 0$, these inequalities define a closed cone with nonempty interior in a space tangent to $\mathcal{V}$. The symmetry group of this cone is the product of the dilatation group $\mathbf{R}^*$ and a subgroup of $O(2, 6)$ which leaves the time-like four-vector $u$ invariant and is isomorphic to $O^{\uparrow}(1, 6)$. It is easier to work in a reference frame in which $\mathbf{u} = 0$, which is a rest frame of the particle if the constraint (\ref{Constr2}) is satisfied. In this frame the condition (\ref{Cond2}) takes the simple form
\begin{equation} \label{Cond3}
\|d \mathbf{x}\|^2 + a_M^{-2} \|d\mathbf{u}\|^2 \leq (dx^0)^2.
\end{equation}
We have obtained a formalism that agrees with the ideas proposed in the preceding section. 

If we also impose the constraint (\ref{Constr2}), the condition (\ref{Cond2}) takes the form
\begin{equation} \label{Cond4}
a_M^{-2} \left(\|d\mathbf{u}\|^2 - (u^0)^{-2} (\mathbf{u} \cdot d \mathbf{u})^2\right) \leq \left(1 - (u^0)^{-2} \|\mathbf{u}\|^2\right)(dx^0)^2,
\end{equation}
or in the rest frame
\begin{equation} \label{Cond5}
a_M^{-2} \|d\mathbf{u}\|^2 \leq (dx^0)^2,
\end{equation}
which implies the upper bound
\begin{equation} \label{Bound}
\|\mathbf{a}\| = \left\|\frac{d\mathbf{u}}{d\tau}\right\| \leq a_M
\end{equation}
for the  acceleration measured in the rest frame. Eq.\ (\ref{Cond4}) and the constraint (\ref{Constr2}) define in the spaces tangent to $\mathcal{V}$ a closed cone with empty interior symmetric with respect to a group isomorphic to 
$O^{\uparrow}(1, 3) \times \mathbf{R}^*$ (not to be interpreted as the usual Lorentz group). 

\section{The Born-Infeld kinematics.}

In this section we consider again the space $\mathcal{W} = \mathbf{R}^8$ with coordinates $x^k$ and $u^k$. We indicate by $W$ a tangent space of $\mathcal{W}$ and, guided by the considerations given in section 1, we try to describe the maximal acceleration by means of a cone $W^+ \subset W$ with the following properties:
\begin{enumerate}
\item It is a closed cone with nonempty interior;
\item It is invariant under the proper orthochronous Lorentz group;
\item It is invariant under the transformation  $dx \to dx$, 
$du \to -du$.
\end{enumerate}
The last property requires a symmetry (indicated by $C$ in the following) with respect to a change of the acceleration sign. We show in the appendix B that these conditions determine $W^+$ up to the choice of the parameter $a_M$ and up to a time inversion.  It is described by the conditions
\begin{equation} \label{Cond6}
dx_+ = dx + a_M^{-1} \, du \in V^+, \qquad dx_- = dx - a_M^{-1} \, du \in V^+.
\end{equation}

From the equality
\begin{equation}
dx_+ \cdot dx_+ + dx_- \cdot dx_-  = 2 d\sigma^2,
\end{equation} 
we see that the condition (\ref{Cond6}) is stronger than eq.\ (\ref{Cond1}). The symmetry group of $W^+$ is $C \times_s (O^{\uparrow}(1, 3) \times \mathbf{R}^* \times O^{\uparrow}(1, 3)  \times \mathbf{R}^*$), where the Lorentz groups and the dilatation groups $\mathbf{R}^*$ act separately on the four-vectors $x_+$ and $x_-$. The reflection $C$, introduced above, exchanges $x_+$ and $x_-$.

As in the preceding section, we have to take into account the constraints (\ref{Constr1}) and (\ref{Constr2}). They are not invariant under the symmetry group, but their consequence (\ref{Constr3}) is invariant, as we see from the equality
\begin{equation}
dx_+ \cdot dx_+ - dx_- \cdot dx_-  = 4a_M^{-1}  dx \cdot du.
\end{equation} 

If we take into account the holonomous constraint (\ref{Constr1}), we obtain a closed cone in a space tangent to $\mathcal{V}$ given by   
\begin{equation} 
\|d\mathbf{x} \pm a_M^{-1} d \mathbf{u}\| \leq dx^0 \pm a_M^{-1} (u^0)^{-1} \mathbf{u} \cdot d \mathbf{u}.
\end{equation}
In a frame in which $\mathbf{u} = 0$ it takes the form
\begin{equation} 
\|d\mathbf{x} \pm a_M^{-1} d \mathbf{u}\| \leq dx^0,
\end{equation}
namely
\begin{equation} \label{Cond7}
\|d\mathbf{x}\|^2 + 2 a_M^{-1} |d \mathbf{x} \cdot d \mathbf{u}| + a_M^{-2} \|d \mathbf{u}\|^2 \leq (dx^0)^2,
\qquad dx^0 \geq 0.
\end{equation}
Note that this cone is smaller than the one defined by eq.\ (\ref{Cond3}). Its symmetry group is $C \times_s (O(3) \times O(3)) \times \mathbf{R}^*$.

If we also take into account the constraint (\ref{Constr2}), we obtain exactly the condition (\ref{Cond4}) or, in the rest frame, the condition (\ref{Cond5}). It follows that, as far as the motion of point particles is concerned, the formalism of the present section is equivalent to the formalism of the preceding section. However, it presents some advantages: the geometry of $\mathcal{W}$ is described by a cone $W^+$, in agreement with the ideas of section 1, and the constraint (\ref{Constr3}) has the same symmetry as the cone.

The formalism described above, and in particular the symmetry group $O^{\uparrow}(1, 3) \times O^{\uparrow}(1, 3)$, coincides with some principles of the ``Born-Infeld kinematics'' proposed in refs.\ \cite{Schuller,Schuller1,SWG,SP,Schuller2} and motivated by the Born-Infeld theory of electrodynamics \cite{BI}. Our considerations may provide an alternative partial justification of those ideas.

In refs.\ \cite{Schuller,Schuller1,SWG,SP,Schuller2} one finds a very elegant treatment obtained by considering the space $W$ as a four-dimensional free module over the ring of the pseudo-complex numbers, which is generated by the real field and by a pseudo-imaginary unit $I$ with the property $I^2 = 1$. We refer to the original papers for more details and for dynamical considerations. Some further comments are given at the end of the next section and in the last section.

\section{The space of reference frames.}

Another approach to the geometry of maximal acceleration is based on the principal fibre bundle of the Lorentz frames, which we indicate by $\mathcal{S}$.  A point of $\mathcal{S}$ corresponds to a tetrad of four-vectors $e_0, \ldots, e_3$ in a tangent space of space-time with the properties
\begin{equation}
e_i \cdot e_k = \eta_{ik}, 
\end{equation} 
where in the right hand side the usual diagonal metric tensor of special relativity appears. We assume that $e_0 \in V^+$ and that the other three four-vectors define a left-handed spatial frame

As in the preceding sections, we disregard gravitation; the extension to the general case does not present difficulties, it is sufficient to replace the infinitesimal translations by parallel displacements. Then we may choose a frame $s_0 \in \mathcal{S}$ and all the other frames have the form $s = gs_0$, where $g$ is an element of the proper orthochronous Poincar\'e group.  In this way we can identify $\mathcal{S}$ with the Poincar\'e group.

An infinitesimal displacement of an element of $\mathcal{S}$ can be obtained by means of an infinitesimal Poincar\'e transformation. This means that the tangent spaces of $\mathcal{S}$ can be identified with the Poincar\'e Lie algebra $\mathcal{T}$. We introduce in $\mathcal{T}$ a basis formed by the vectors $A_i$ and $A_{ik} = -A_{ki}$ which represent, respectively, the generators of the space-time translations and of the Lorentz transformations. Under the Lorentz group, they transform as the components of a four-vector and of an antisymmetric tensor. They form a set of ten vector fields in the manifold $\mathcal{S}$.

The Hamiltonian dynamics of a particle in the space $\mathcal{S}$ has been discussed in ref.\ \cite{Kunzle}.  As in the preceding sections, we only consider the kinematical aspects. To the particle motion we associate a trajectory in $\mathcal{S}$ given by $\tau \to s(\tau) \in \mathcal{S}$ and we write
\begin{equation} \label{Deriv}
\frac{ds(\tau)}{d \tau} = b^{i} A_{i} + 2^{-1} b^{ik} A_{ik} 
\in \mathcal{T}, \qquad b^{ik} = - b^{ki}.
\end{equation} 
We also introduce the vectors
\begin{equation}
\mathbf{b} =(b^1, b^2, b^3), \qquad
\mathbf{b}' =(b^{12}, b^{23}, b^{31}), \qquad
\mathbf{b}'' =(b^{10}, b^{20}, b^{30}),
\end{equation}
which describe space translations, rotations and Lorentz boosts.

The analogy argument presented in Section 1 suggests that the maximal acceleration hypothesis can be formulated by requiring that the vector (\ref{Deriv}) belongs to a cone $\mathcal{T}^+ \subset  \mathcal{T}$ with the properties
\begin{enumerate}
\item $\mathcal{T}^+$ is a closed cone with nonempty interior;
\item It is invariant under the proper orthochronous Lorentz group.
\end{enumerate}
It has been proved in refs.\ \cite{Toller1,Toller2} that these properties determine $\mathcal{T}^+$ up to the value of the parameter $a_M$ and up to an inversion of all the coordinates. In the following we choose the units in such a way that $a_M = c = 1$.

A simple definition of $\mathcal{T}^+$ is based on a representation of the elements of $\mathcal{T}$ by means of $4 \times 4$ matrices given by
\begin{equation}
\hat b = b^i \gamma_0 \gamma_i
+ i 2^{-1} b^{ik} \gamma_0 \gamma_{i} \gamma_{k},
\end{equation} 
where $\gamma_{i}$ are the Dirac matrices in the Majorana representation. One can show that the matrix $\hat b$ is real and symmetric and one defines $\mathcal{T}^+$ by requiring that it is positive semi-definite, namely that
\begin{equation} \label{Cond8}
\psi^T \hat b \psi \geq 0
\end{equation}
for any choice of the real spinor $\psi$. This condition implies the inequalities
\begin{equation}
\|\mathbf{b}\| \leq b^0, \qquad \|\mathbf{b}'\| \leq b^0, \qquad \|\mathbf{b}''\| \leq b^0, 
\end{equation}
but it is stronger than them.

It is clear that the cone $\mathcal{T}^+$ has the symmetry group $GL(4, \mathbf{R})$ acting as
\begin{equation} 
\hat b \to a \hat b a^T, \qquad a \in GL(4, \mathbf{R}).
\end{equation}
This group represents a broken symmetry, since the structure constants of the Poincar\'e algebra, considered as external fields that  describe the vacuum, are not invariant \cite{Toller1}. It may be interesting to recall that the special linear group $SL(4, \mathbf{R})$ is locally isomorphic to the pseudo-orthogonal group $SO(3, 3)$.

As in the approaches of the preceding sections, a line which represents the motion of a particle must satisfy some constraints. However, there are no holonomous constraints of the kind (\ref{Constr1}) and all the points of $\mathcal{S}$ are physically relevant. We require that $s(\tau)$ is a rest frame of the particle. It follows that the four-vector $e_0$ is just the four-velocity $u$ introduced in the preceding sections and that
\begin{equation} \label{Constr4}
b^0 = 1, \qquad \mathbf{b} = 0.
\end{equation} 
Moreover, we require that the frame $s(\tau)$ is accelerated without rotation, namely it is Fermi-Walker transported. In this way we obtain the constraint
\begin{equation} \label{Constr5}
\mathbf{b}'  = 0.
\end{equation}

We remain with the components of the acceleration
\begin{equation} 
\mathbf{a} = (b^0)^{-1} \mathbf{b}'',
\end{equation}
measured in the rest frame $s(\tau)$. Our condition means that the matrix
\begin{equation} 
1 + i \mathbf{a} \cdot \boldsymbol{\gamma}, \qquad  
\boldsymbol{\gamma} = (\gamma^1, \gamma^2, \gamma^3),
\end{equation}
is positive semi-definite and it follows that the acceleration has the upper bound (\ref{Bound}).

If we consider a particle with spin or an extended particle, the cone $\mathcal{T}^+$ contains some more information. In fact, in this case it is natural to give up the constraint (\ref{Constr5}) and to assume that the frame $s(\tau)$ rotates together with the particle with an angular velocity
\begin{equation} 
\boldsymbol{\omega} = (b^0)^{-1} \mathbf{b}'
\end{equation}
with respect to a Fermi-Walker transported frame. It has been shown in ref.\ \cite{Toller2} that from our condition one gets the upper bound
\begin{equation} \label{Omega}
\|\boldsymbol{\omega}\|^2 + \|\mathbf{a}\|^2 + 2 \| \mathbf{a} \times \boldsymbol{\omega} \| \leq a_M^2.
\end{equation} 
The set defined by this inequality can be considered as the intersection of the cone $\mathcal{T}^+$ with the plane defined by eq.\ (\ref{Constr4}). It is convex and compact, but not Lorentz invariant, as it deals with quantities measured in the rest frame.

In ref.\ \cite{Toller2} one also finds a justification of this formula in terms of a model that considers a spherical rigid body with radius $a_M^{-1}$, requiring that all its points have a velocity smaller than $c$. This formula implies, for instance, an upper bound to the precession angular velocity of a particle with a magnetic moment in a magnetic field. 

In refs.\ \cite{SP,Schuller2} the Born-Infeld kinematics is also treated from the point of view of the bundle of frames.  We cannot give here a consistent summary of this development, which is essentially based on the pseudo-complex geometry and on the Born-Infeld dynamics, but some results can be expressed with the notations introduced above and it is interesting to compare them with the point of view described in the present section.  

The relevant formula is
\begin{equation} 
b^i{}_k = a_M b^0 [\tanh \epsilon]^i{}_k,
\end{equation} 
where the matrix $\epsilon$ has the antisymmetry property
\begin{equation} 
\eta_{ij} \epsilon^j{}_k + \eta_{kj} \epsilon^j{}_i = 0,
\end{equation} 
namely it represents an element of the Lorentz Lie algebra $o(1, 3)$. It can be expressed in terms of the forces acting on the particle.

The function $\tanh$ is defined by a power series of matrices and, as in the real numeric case, its range is not the whole vector space $o(1, 3)$.  The inequalities that describe this range can be calculated by choosing a suitable Lorentz frame \cite{SP,Schuller2} and one obtains
\begin{equation} 
\|\mathbf{a}\|^2 - \|\boldsymbol{\omega}\|^2 + 
a_M^{-2} (\mathbf{a} \cdot \boldsymbol{\omega})^2 \leq a_M^2.
\end{equation} 
This inequality is essentially different from eq.\ (\ref{Omega}), in particular it does not define a convex set, it allows any value of the angular velocity $\boldsymbol{\omega}$ and it is Lorentz invariant. If one is dealing with a point particle, one can put $\boldsymbol{\omega} = 0$ and the two formulas coincide.

\section{Feasible transformations.}

The treatment of the preceding section, even after the introduction of rotating particles, still maintains the constraint (\ref{Constr4}), which does not permit us to give a physical meaning to all the vectors belonging to $\mathcal{T}^+$. One can overcome this obstacle by adopting a different interpretation, which is the one originally proposed in ref.\ \cite{Toller1} and is based on the ideas discussed in ref.\  \cite{Toller3}. 

The starting point is the remark that a local reference frame is defined by some material object and any physical procedure, in order to be localized in space and time, must refer to some of these objects. In particular, a procedure which has the aim to build a new (material) reference frame starting from a pre-existent one is called a {\it transformation}.  

A transformation is described, at least approximately, by an element of the Poincar\'e group, but only some elements correspond to {\it feasible} transformations, because, as pointed out in ref.\  \cite{Toller3}, it takes some time to translate, accelerate or rotate a (material) reference frame.  The infinitesimal feasible transformations are described by elements of the cone $\mathcal{T}^+ \subset  \mathcal{T}$  described in the preceding section. The assumption that  $\mathcal{T}^+$ is a cone means that the composition of two feasible infinitesimal transformations is again feasible.

This point of view can also be applied to the space-time-velocity manifold $\mathcal{V}$ and it permits a physical interpretation of tangent vectors that do not respect the constraint (\ref{Constr2}). A point of $\mathcal{V}$ represents the position of the origin and the time-like four-vector $e_0 = u$ of a tetrad, without giving any information about the three other spacelike four-vectors. However, $u$ is not interpreted as the four-velocity of a particle and the constraint (\ref{Constr2}) is not necessary.  One may say that a point of $\mathcal{V}$ represents a class of reference frames which are (partially) physically defined by a spherically symmetric material object.

In this way we have defined a mapping $\mathcal{S} \to \mathcal{V}$. Points of $\mathcal{S}$ which differ by a rotation correspond to the same point of $\mathcal{V}$ and we can represent $\mathcal{V}$ as a quotient space
\begin{equation}  \label{Quotient}
\mathcal{V} = \mathcal{S}/SO(3).
\end{equation}
This formula provides a bridge between the formalisms of the sections 2, 3 and the one of section 4.

The corresponding linear mapping between a tangent space of $\mathcal{S}$, identified with $\mathcal{T}$, and a tangent space of $\mathcal{V}$ introduces in the latter tangent space the seven coordinates $b^0, \mathbf{b}, \mathbf{b}''$ (defined up to a rotation). This mapping transforms the cone $\mathcal{T}^+$ into a cone tangent to $\mathcal{V}$. It is interestig to compare this cone with the cones (\ref{Cond3}) and (\ref{Cond7}) defined in the section 2 and 3, which, with the new notation, take, respectively, the form
\begin{equation} \label{Cond9}
\|\mathbf{b}\|^2 + \|\mathbf{b}''\|^2 \leq (b^0)^2,
\qquad b^0 \geq 0,
\end{equation}
\begin{equation} \label{Cond10}
\|\mathbf{b}\|^2 + 2 |\mathbf{b} \cdot \mathbf{b}''| + \|\mathbf{b}''\|^2 \leq (b^0)^2,
\qquad b^0 \geq 0.
\end{equation}

We start with another description of $\mathcal{T}^+$ as the cone generated by the elements of the form \cite{Toller1,Toller2}
\begin{equation} \label{Extr}
\|\mathbf{b}\| = \|\mathbf{b}''\| = b^0, \qquad  \mathbf{b} \cdot \mathbf{b}'' = 0,
\end{equation}
\begin{equation} 
\mathbf{b}'  =  (b^0)^{-1} \mathbf{b}'' \times \mathbf{b}.
\end{equation}
These equations define the extremal elements of $\mathcal{T}^+$. Their projections on a tangent space of $\mathcal{V}$ are defined by eq.\ (\ref{Extr}) alone and they generate the projection of the cone $\mathcal{T}^+$ in this space. The elements defined by eq.\ (\ref{Extr}) satisfy neither eq.\ (\ref{Cond9}) nor eq.\ (\ref{Cond10}) and they generate a cone different from the cones defined by these equations. 

\section{Conclusions and outlook.}

In sections 2, 3, and 4 we have discussed three different geometric descriptions of the maximal acceleration hypothesis. The first and the second are based on the relativistic phase space $\mathcal{W}$ and have, respectively, the symmetry groups $O(2, 6) \times \mathbf{R}^*$ and $C \times_s (O^{\uparrow}(1, 3) \times \mathbf{R}^* \times O^{\uparrow}(1, 3) \times \mathbf{R}^*)$, while the third is based on the bundle $\mathcal{S}$ of the Lorentz frames and has the symmetry group $GL(4, \mathbf{R})$, acting linearly on Majorana spinors (not on the space-time coordinates). The third formalism also allows a treatment of the rotational degrees of freedom and implies an upper bound to the angular velocity.

Note that the above mentioned symmetry groups concern some aspects of particle kinematics, while other aspects of kinematics and dynamics have lower symmetry properties unless they are modified. Since the action of these groups mixes the space-time coordinates with other variables, the important concept of space-time coincidence of two events looses the absolute character it has in relativity theory. 

In order to compare the three formalisms, one has to introduce the space-time-velocity manifold $\mathcal{V}$, obtained from $\mathcal{W}$ by imposing the ``on shell'' constraint (\ref{Constr1}), or from the space $\mathcal{S}$ by identifying the frames that differ for a rotation, namely by performing the quotient (\ref{Quotient}).  We have seen that the three approaches define three different cones in the tangent spaces of $\mathcal{V}$.
 
However, the variables that describe a point particle are subject to additional constraints and when these constraints are taken into account the three formalisms are equivalent. In order to give a physical meaning to the differences we have found, we have to go beyond the particle interpretation. An interpretation in terms of local reference frames and infinitesimal transformations is summarized in section 5. From this wider point of view, the three geometries are physically  different. We have also seen that other differences appear when one considers rotating particles.

Since experimental tests look rather difficult, the choice between these kinematic descriptions depends on the success of the attempts to find an underlying dynamical scheme in which the limitations to the acceleration appear naturally and not as an extraneous requirement superimposed to a preceding theory. Some general remarks will be given in a forthcoming paper.

Finally, we show that, if the formalism of section 3 is interpreted as an upper bound $f_M$ to the force (instead of the acceleration), it is compatible with the formalism of section 4 and one can apply both the ideas at the same time. This can be obtained by considering the cotangent bundle $T^*\mathcal{S}$ of the manifold $\mathcal{S}$ and introducing in each cotangent space the coordinates $p_k$ and $p_{ik} = - p_{ki}$, which represent, respectively, the four-momentum and the relativistic angular momentum, as it is discussed in detail in refs. \cite{Toller4,Vanzo}.

In order to simplify the notation, it is convenient to introduce the greek indices $\alpha, \beta$, which take the values $0, 1 ,\ldots, 9$ and to indicate the quantities introduced above by $p_{\alpha}$. In a similar way we indicate the quantities $b^i$ and $b^{ik}$ by $b^{\alpha}$. There is no $GL(4, \mathbf{R})$ invariant metric in the ten-dimensional space $\mathcal{T}$, which permits one to raise or lower the greek indices. However, as we shall discuss elsewhere, one can choose a subgroup of $GL(4, \mathbf{R})$ isomorphic to the symplectic group $Sp(4, \mathbf{R})$ and locally isomorphic to the  anti-de Sitter group $SO(2, 3)$, which is still sufficient to fix the value of $a_M$ and admits an invariant metric tensor. By means of this tensor one can define the quantities $p^{\alpha}$ that transform under $Sp(4, \mathbf{R})$ in the same way as $b^{\alpha}$.  We indicate by $\dot p^{\alpha}$ the derivative with respect to the parameter $\tau$ which appears in eq.\ (\ref{Deriv}).

The ideas of sections 3 and 4 can be condensated in the inequalities
\begin{equation} \label{Cond11}
(b_+^{\alpha}) = (b^{\alpha} + f_M^{-1} \dot p^{\alpha}) \in \mathcal{T}^+, \qquad
(b_-^{\alpha}) = (b^{\alpha} - f_M^{-1} \dot p^{\alpha}) \in \mathcal{T}^+
\end{equation}
which generalize eq.\ (\ref{Cond6}). They define in the tangent spaces of $T^*\mathcal{S}$ a cone symmetric with respect to the group $GL(4, \mathbf{R}) \times GL(4, \mathbf{R})$, though only the subgroup $Sp(4, \mathbf{R}) \times Sp(4, \mathbf{R})$ is physically relevant. This symmetry group contains the subgroup $SL(2, \mathbf{C)} \times SL(2, \mathbf{C)}$, which is the universal covering of the group $SO^{\uparrow}(1, 3) \times  SO^{\uparrow}(1, 3)$ introduced in section 3, and also the diagonal subgroup $GL(4, \mathbf{R})$ introduced in section 4.

The cone (\ref{Cond11}) describes upper bounds to velocity, angular velocity, acceleration, power, force and torque. The corresponding geometry contains the fundamental constants $c$, $a_M$ and $f_M$ and therefore it fixes the fundamental scales for all the physical quantities, in particular for the action. A classical (non quantum) treatment is justified only if 
\begin{equation} 
f_M a_M^{-2} c^3 \gg \hbar.
\end{equation} 

\appendix
\renewcommand{\thesection}{Appendix \Alph{section}:}

\section{Characterization of the cone $V^+$.}

As an introduction to the less trivial proof of the next appendix,  we show that the conditions a) and b) of section 1 define the cone $V^+$ up to the choice of the time and length units and up to a time inversion. 

First we remark that the intersection of the cone and the three-dimensional plane $x^0 = \tau$ is a closed convex rotation invariant set, namely the whole plane, a closed ball, a point or the empty set. The whole plane must be excluded, because it contains straight lines. Since the cone cannot contain the straight line $\mathbf{x} = 0$, the intersection must be empty for $\tau < 0$ or for $\tau > 0$. The second possibility is reduced to the first one by means of a time inversion.

The intersection with the plane $x^0 = 1$ cannot be reduced to a point, otherwise the cone is reduced to a half line, which has an empty interior.  We choose the unit of lenght in such a way that it is a ball with unit radius. The cone is the union of the half lines starting from the origin that intersect this ball.

\section{Characterization of the cone $W^+$.}

In this appendix we show that the conditions a), b) and c) of section 3 define the cone $W^+$ described by eq.\ (\ref{Cond6}) with a suitable choice of the parameter $a_M$ and up to a time inversion. 

To indicate an element of $W^+$, we use the simpler notation $(x, u)$ (instead of $(dx, du)$), where $x$ and $u$ are four-vectors. From the assumption c) and the convexity property we see that if  $(x, u) \in W^+$ we have also $(x, -u) \in W^+$ and $(x, 0) \in W^+$. The last condition means that $x$ belongs to a Lorentz-invariant closed cone in the four-vector space, namely $V^+$ or $-V^+$. Then, possibly after a time inversion, we have $x \in V^+$.

We choose $k \in L^+$, namely a light-like four-vector $k$ with $k^0 > 0$ and we consider the equations
\begin{equation}
y^0 = k \cdot x \geq 0, \qquad y^1 = k \cdot u,
\end{equation}
which define a linear mapping $W \to \mathbf{R}^2$. The image $I$ of this mapping is a wedge in the two-dimensional half-plane $y^0 \geq 0$, invariant under the transformation $y^1 \to - y^1$. Any pair of four-vectors $k, k' \in L^+$  are connected by an orhochronous Lorentz transformation. Since the cone $W^+$ is invariant with respect to these transformations, we see that the image $I$ does not depend on the particular choice of $k$.

We indicate by $(x', u')$ the element obtained from $(x, u) \in W^+$ by means of a Lorentz boost with rapidity $\zeta$ in the direction of $-\mathbf{k}$, the multiplication by the factor $2\exp(-\zeta)$ and the limit $\zeta \to +\infty$. Since the cone $W^+$ is Lorentz invariant and closed, $(x', u')$ belongs to $W^+$.  In a suitable reference frame we have
\begin{equation}
k = (1, 0, 0, -1), \qquad y^0 = x^0 + x^3, \qquad y^1 = u^0 + u^3,
\end{equation}
\begin{eqnarray}
&x' = \lim_{\zeta \to +\infty}2\exp(-\zeta)(x^0 \cosh\zeta + x^3 \sinh\zeta, x^1, x^2,  x^0\sinh\zeta + x^3 \cosh\zeta) =&
\nonumber \\ 
&= (x^0 + x^3, 0, 0, x^0 + x^3) = y^0 h,&
\end{eqnarray}
\begin{eqnarray}
&u' = \lim_{\zeta \to +\infty}2\exp(-\zeta)(u^0 \cosh\zeta + u^3 \sinh\zeta, u^1, u^2,  u^0\sinh\zeta + u^3 \cosh\zeta) =& 
\nonumber \\ 
&= (u^0 + u^3, 0, 0, u^0 + u^3) = y^1 h,&
\end{eqnarray}
where
\begin{equation}
h = (1, 0, 0, 1).
\end{equation}

We have seen that $(y^0, y^1) \in I$, if and only if $(y^0 h, y^1 h) \in W^+$. We have proven this property for a particular choice of the four-vector $h$, but it follows from the Lorentz symmetry that it holds for any $h \in L^+$.  It follows that $I$, being the inverse image of $W^+$ under the continuous mapping $(y^0, y^1) \to (y^0 h, y^1 h)$, is closed. 

If $I$ is the whole closed half plane, one can easily see that $W^+$ contains the plane containing the elements $(0, u)$, in contradiction with our assumptions. Then, $I$ is a closed cone defined by
\begin{equation}
|y^1| \leq a_M y^0,
\end{equation}
where $a_M$ is a positive number.
This means that $(x, u) \in W^+$ implies
\begin{equation}
k \cdot (a_M x \pm u) \geq 0
\end{equation}
for all the four-vectors $k \in L^+$. This condition is equivalent to eq.\ (\ref{Cond6}). On the other hand, $W^+$ contains all the elements of the kind $(h, \pm a_M h)$ with $h \in L^+$. In other words, it contains all the elements with $x_+ = 2h, \,\, x_- = 0$ and with $x_+ = 0, \,\, x_- = 2h$ and by convexity all the elements which satisfy the condition (\ref{Cond6}).

\end{document}